# Radiocaesium Activity Concentrations in Potatoes in Croatia after the Chernobyl Accident and Dose Assessment

Zdenko Franić[1*], Branko Petrinec[1], Gordana Marović[1] and Zrinka Franić[2]

[1] Radiation Protection Unit, Institute for Medical Research and Occupational Health, Zagreb, Croatia

[2] Medical School, University of Zagreb, Zagreb, Croatia

**ABSTRACT**

Systematic investigations of $^{137}$Cs and $^{134}$Cs activity concentrations in potatoes (*Solanum tuberosum*) for the post-Chernobyl period (1986-2005) in the Republic of Croatia are summarized. The correlation between $^{137}$Cs activity concentrations in fallout and potatoes, has been found to be very good, the correlation coefficient being $r^2=0.88$ with $P(t) < 0.001$ for 18 degrees of freedom. As the radiocaesium levels in potatoes decreased exponentially, the mean residence time of $^{137}$Cs in potatoes was estimated by fitting the measured activity concentrations to the exponential curve. The mean residence time was found to be $6.8 \pm 1.1$ years, the standard deviation being estimated by the Monte Carlo simulations.

The initial observed $^{134}$Cs:$^{137}$Cs activity ratio in potatoes has been found to be quite variable, but slightly lesser than theoretically predicted value of 0.5, calculated by applying the known inventory of these radionuclides in the Chernobyl reactor to the equation for the differential radioactive decay. This can be explained by presence of the pre-Chernobyl $^{137}$Cs in soil that originated from nuclear fallout.

The annual effective doses received by $^{134}$Cs and $^{137}$Cs intake due to consumption of potatoes estimated for an adult member of Croatian population were found to be very small, as the per caput dose for the entire 1986 – 2005 period was calculated to be about 2.9 μSv, $^{134}$Cs accounting approximately for 1/3 of the entire dose. Therefore, after the Chernobyl accident consumption of potatoes was not the critical pathway for human intake of radiocaesium from the environment in Croatia.

------------------------

* Address correspondence to Dr. Zdenko Franić, Radiation Protection Unit, Institute for Medical Research and Occupational Health, PO Box 291, HR-10000 Zagreb, Croatia,
Phone: +385-1-4673188 ext: 132, Fax: +385-1-4673303; E-mail: franić@imi.hr

INTRODUCTION

Tests of nuclear weapons conducted in the atmosphere and releases of radioactive material from nuclear facilities are the main causes of the man-made radioactive contamination of human environment. Once released to the atmosphere, long-range atmospheric transport processes can cause a widespread distribution of such radioactive matter, although it may, like in the case of Chernobyl accident, originate in a single point on the Earth's surface. The resulting fallout, consisting of short and long-lived radionuclides, eventually affects humans, either directly or indirectly by entering the food chain through consumption of plants and animals. In both cases fallout causes a health hazard to the population through the direct irradiation and consumption of contaminated foodstuffs.

Due to a comparatively high contribution of the ingestion doses to the total dose received by population after the exposure to nuclear fallout, a reliable prognosis of the expected ingestion doses is of utmost importance. The ingestion dose strongly depends on the consumption of various types of foodstuffs, and related activity concentrations of respective radionuclides in those foodstuffs, which themselves usually depend upon the transfer from fallout. In addition, a reliable prediction of the expected ingestion dose received by consumption of a particular foodstuff requires the detailed knowledge of decreasing behaviour of activity concentrations in the environment and respective foodstuffs.

Among man-made radioactive nuclides, those of the radiocaesium, particularly $^{137}$Cs, are regarded as a great potential hazard to living beings. Namely, this fission product has unique combination of relatively long half-lives (30.14 years) and chemical and metabolic properties resembling those of the potassium, which is principal constituent of cell tissue.

Investigations of the distribution and fate of natural, nuclear weapons produced and reactor-released radionuclides in foodstuffs have been conducted as a part of an extended and still ongoing monitoring programme of radioactive contamination of human environment in Croatia. (Popović at al. 1966 – 1978, Bauman et al. 1979 – 1992, Kovač et al. 1993 – 1998 and Marović et al. 1999 – 2005). However, the regular investigations of radioactive contamination of potatoes (*Solanum tuberosum*) have started in the year of the Chernobyl accident, i.e. in 1986.

The annual production of potatoes in Croatia is about 550,000 tonnes annually (Central Bureau of Statistics, 2004). Approximately 1/3 of this quantity is being consumed in Croatia. Since the population of Croatia is about 4.5 million of inhabitants, the consumption rate of potatoes is approximately 40 kg per person annually. Therefore it can be regarded as quite important foodstuff in Croatia which can have, if contaminated, severe radiological consequences for the entire population.

**MATERIALS AND METHODS**

Potatoes have been obtained commercially on civil markets in the cities of Zagreb (45° 50' N, 16° 00' E), Osijek (45° 30' N, 18° 40' E) and Zadar (44° 06' N, 15° 15' E). From each site were obtained several kilograms of potatoes that were first peeled of and then cut into small pieces in order to obtain the composite sample. Samples were dried in an oven and then ashed in a muffle furnace at 450 °C for 24 h.

Fallout samples were collected monthly in the city of Zagreb at the location of the Institute for Medical Research and Occupational Health (45° 50' 07.3" N, 15° 58' 58.7" E). The funnels, which were used for rainwater collection, had a 1 m$^2$ catchment area. Precipitation height was measured by Hellman pluviometer. Rainwater was evaporated to volume of 1 L in order to enrich the $^{137}$Cs activity concentration.

A gamma-ray spectrometry system based on a low-level ORTEC Ge(Li) detector (FWHM 1.87 keV at 1.33 MeV $^{60}$Co and relative efficacy of 15.4% at 1.33 MeV), coupled to a computerized data acquisition system, was used to determine radiocaesium levels in the samples from their gamma-ray spectra. Ash from the samples was measured in cylindrical plastic containers of appropriate volume, which were placed directly on the detector. Fallout samples were measured in Marinelli beakers. Counting times, which depend on radiocaesium activity concentrations in samples, ranged from 10,000 to 250,000 seconds and typically were 80,000 s.

Quality assurance and intercalibration measurements were performed through participation in an International Atomic Energy Agency (IAEA) and World Health Organization (WHO) international intercalibration programmes, which also include the regular performance of blank, background and quality control measurements.

Radiocaesium activity concentrations in samples in this paper are reported as averages of three sampling locations, implicitly implying similar characteristics and microclimate conditions, of soils in which potatoes have been grown, that is not necessarily true. However, typical relative error was about 25 %. It should be noted that similar approach has been implemented in the radioecological investigations of contamination of diet components in Nordic countries (Strandberg, 1994 and Aarkrog, 1994).

## RESULTS AND DISCUSSION

*$^{137}$Cs and $^{134}$C activity concentrations in potatoes*

The radioactive fallout resulting from large-scale nuclear weapon tests in the atmosphere conducted in the 1960s, followed by similar, but smaller scale tests by the Chinese and French in the 1970s and afterwards, was the dominant route for the introduction of artificial radionuclides in the environment until the nuclear accident at Chernobyl, in former USSR, on 26 April 1986. Fortunately, due to the prevailing meteorological conditions at the time after the accident that influenced the formation and spreading direction of Chernobyl plumes, Croatia was only on the North-Western region partially affected by the edge of one of the plumes (UNSCEAR, 1988).

The estimated amount of radiocaesium released after the reactor explosion at Chernobyl was $3.7 \times 10^{16}$ Bq of $^{137}$Cs (13% of total reactor inventory) and $1.9 \times 10^{16}$ Bq of $^{134}$Cs (10% of total reactor inventory (International Atomic Energy Agency, 1986). Thus, the initial value for the $^{134}$Cs:$^{137}$Cs activity ratio in May 1986 was 0.51.

The highest observed $^{137}$Cs activity concentrations in fallout were recorded in May 1986, resulting in 6410 Bq m$^{-2}$ for the surface deposit by fallout in the year 1986 (Bauman et al. 1979 – 1992; Franić 1992). The radioactive material introduced to the atmosphere by Chernobyl accident was by global dispersion processes distributed throughout the troposphere, causing the increased radiocaesium activity concentrations in the environment in years to come. However, $^{137}$Cs showed a significant exponential decrease over time because of natural removal as well as radioactive decay. Also, no new releases of $^{137}$Cs occurred after the Chernobyl reactor accident either from nuclear facilities or nuclear weapons testing. Therefore, in 2005 total $^{137}$Cs surface deposit by fallout was only 2.8 Bq m$^{-2}$.

The highest $^{137}$Cs activity concentration in potatoes, being $1.100 \pm 0.650$ Bq kg$^{-1}$ was recorded in 1986, while in 2001 was recorded minimal value of only $0.033 \pm 0.011$ Bq kg$^{-1}$, the $^{137}$Cs activity concentration in 2005 being $0.055 \pm 0.023$ Bq kg$^{-1}$. The transient increase of $^{137}$Cs activity concentration in year 1990 could be explained by penetration of caesium to deeper soil layers as root uptake by plants of radionuclides as well as other stable elements depends upon their vertical distribution. It should be noted that these activity concentrations are comparable to the values observed elsewhere (Tsukada and Nakamura 1999, Schwaiger et al. 2004).

When discussing $^{134}$Cs activity concentrations, it should be noted that $^{134}$Cs and $^{137}$Cs, being the most conservative in behaviour, have undergone no selective removal in transit between the accident site at Chernobyl and Croatia as their activity ratio value of 0.5, found in various samples taken on the accident site, has not been altered as this ratio has been found in most of the environmental samples in Croatia (Franić et al 1988; Franić et al 1992a; Franić et al 2006). However, the initial observed $^{134}$Cs:$^{137}$Cs activity ratio in potatoes has been found to be quite variable, but slightly lesser than 0.5. This can be explained by presence of the pre-Chernobyl $^{137}$Cs in soil that originated from nuclear bomb fallout. Consequently, $^{134}$Cs:$^{137}$Cs activity concentration ratio found in potatoes resembled that found in Chernobyl fallout.

*Mean residence time of $^{137}$Cs in potatoes*

A reliable prediction of the expected ingestion dose received by consumption of a particular foodstuff requires the detailed knowledge of decreasing behaviour of activity concentrations in the environment and respective foodstuffs. To evaluate the decrease activity concentrations and to assess the effective mean residence time (i.e., ecological time) of $^{137}$Cs in potatoes the observed data were fitted to the exponential function:

$$A_{pot}(t) = A_{pot}(0)\, e^{-kt} \qquad /1/$$

where:

$A_{pot}(t)$ is time-dependant activity concentration of $^{137}$Cs in potatoes (Bqkg$^{-1}$),

$A_{pot}(0)$ initial activity concentration of $^{137}$Cs in potatoes (Bqkg$^{-1}$) and

$1/k = T_M$ effective (observed) mean residence time of $^{137}$Cs in potatoes (years).

By fitting the data for $^{137}$Cs activity concentration in potatoes to the function /1/ (r = 0.88 with P(t) < 0.001 for 18 degrees of freedom) the effective mean residence time for $^{137}$Cs in potatoes, for the overall observed post-Chernobyl period (1986 – 2005) was estimated to be 6.8 years. This value is in excellent agreement with effective mean residence time of 6.7

± 1.3 y (effective half-life is 1688 ± 338 days) reported for potatoes in Austria (Schwaiger et al 2004), and similar to mean residence time for the other foodstuffs (Schwaiger et al 2004).

The fit is shown on Figure 1.

FIGURE 1 ABOUT HERE

However, to find the real residence time, $T_R$, the observed constant k should be corrected for the radioactive decay. Therefore constant k from the equation /1/ can be written as:

$$k = \lambda + k_R \qquad /2/$$

where $\ln(2)/\lambda = 30.14$ y is the physical half-life of $^{137}$Cs.

From equation (3) real mean residence time for $^{137}$Cs in potatoes, $T_R = 1/k_R$, was found to be 8.0 y, which is about one year higher compared to the effective mean residence time $T_M$.

In order to obtain the standard deviation of $T_M$, Monte Carlo simulations were performed. To be on a conservative side, as well as to simplify calculations, the uniform distribution has been assumed over the $A \pm 2F$ value of $^{137}$Cs activity concentrations in potatoes for respective years, although normal would be more realistic. For each year the random value was generated over the interval $[A - 2F, A + 2F]$ and then from such set of data was estimated the $1/k$ value by fitting to the equation /1/. The process has been repeated 500 times and 500 values for $1/k = T_M$ were obtained. From such set of data the mean value and standard deviation for $T_M$ were calculated to be 6.8 ± 1.1 years.

*$^{137}$Cs transfer from fallout to potatoes*

The intensity of transfer processes that lead from emission of radionuclides to the biosphere to their concentrations in physical environment, living organisms and finally to humans is dependant on various parameters and factors inherent to the different environments, i.e., compartments of the biosphere. These parameters and factors determine the behaviour of radionuclides and eventually the response of the environment. One of the tools that help to compare sensitivities of various environments to radioactive contamination, at least to some extent, is radioecological sensitivity.

Radioecological sensitivity, *Rs*, is defined as the infinite integral of activity concentrations of particular radionuclide in an environmental sample to the integrated deposition. *Rs* is sometimes also called the transfer coefficient from fallout to sample and in the case of food samples is equivalent to UNSCEAR's transfer coefficient $P_{23}$ (UNSCEAR, 1982). Mathematically, $P_{23}$ is defined as follows:

$$P_{23} = \frac{\int_0^\infty A(t)dt}{\int_0^\infty \dot{U}(t)dt} \qquad /3/$$

where:

*A(t)*     is the activity of given radionuclide in food, i.e., potatoes (Bq kg$^{-1}$) and

$\dot{U}(t)$    the fallout deposition rate of this radionuclide (Bq m$^{-2}$ y$^{-1}$).

As for values of *A(t)* and $\dot{U}(t)$ assessed on the yearly basis the integration can be replaced by summation, for the overall observed period, i.e. from 1986 to 2005, the value of $P_{23}$ for $^{137}$Cs in potatoes can be easily calculated to be $4.85 \times 10^{-4}$ Bq y kg$^{-1}$ / Bq m$^{-2}$. That means that with each Becquerel deposited on an area of one square metre of soil by fallout, the activity of one ton of potatoes increases approximately by 0.5 Bq.

However, as in the year 1986 was very high direct $^{137}$Cs deposition, it affects the overall result. Therefore, when the year 1986 is excluded from analysis, $P_{23}$ for $^{137}$Cs in potatoes for 1987 – 2005 period is calculated to be $1.47 \times 10^{-3}$ Bq y kg$^{-1}$ / Bq m$^{-2}$.

To put the obtained values into perspective, the $^{137}$Cs transfer coefficient $P_{23}$ for total diet was estimated to be approximately $1.2 \times 10^{-2}$ Bq y kg$^{-1}$/(Bq m$^{-2}$) for the 1962 - 1979 period in New York (reference location for a northern hemisphere) and $8.0 \times 10^{-3}$ Bq y kg$^{-1}$ / Bq m$^{-2}$ for the 1963 - 1973 period in Argentina (southern hemisphere) (UNSCEAR, 1982).

Also, for $^{137}$Cs in wheat grains for 1965 to 2003 period, the value of $P_{23}$ was estimated to be $8.59 \times 10^{-3}$ Bq y kg$^{-1}$ / Bq m$^{-2}$ (Franić et al 2006). However, when wheat data for pre-Chernobyl and post Chernobyl periods have been considered separately, $P_{23}$ values of $2.79 \times 10^{-2}$ and $5.11 \times 10^{-3}$ Bq y kg$^{-1}$ / Bq m$^{-2}$ were obtained for 1965 - 1985 and 1986 - 2003 periods respectively.

*Modelling $^{137}$Cs activity concentrations in potatoes*

In 1990 and afterwards, contamination of potatoes by the fallout $^{134}$Cs that originated from the Chernobyl nuclear accident was detectable only at a very low level.

$^{137}$Cs activity concentrations in potatoes are in good correlation with fallout activity when related by a simple linear relation, with coefficient of correlation r = 0.88 with P(t) < 0.001 for 18 degrees of freedom. Thus, for the observed period, i.e. from 1986 – 2005, $^{137}$Cs activity concentrations in potatoes can be from fallout data modelled as:

$$A_{pot}(t) = 0.00015 \times A_{fall}(t) + 0.14078 \qquad /4/$$

where:

$A_{pot}(t)$ is $^{137}$Cs activity concentration in potatoes (Bq kg$^{-1}$) and

$A_{fall}(t)$ $^{137}$Cs fallout activity (Bq m$^{-2}$).

By plotting the observed (measured) $^{137}$Cs activity concentrations data in potatoes and data modelled using the equation /4/ against the time scale, it is clear that the fit overestimates the real data ever since year 1995.

The better fit is obtained when $^{137}$Cs activity concentration data in potatoes are fitted to a power function:

$$A_{pot}(t) = 0.06 \, A_{fall}(t)^{0.317} \qquad /5/$$

For equation /5/ coefficient of correlation between observed and modelled $^{137}$Cs activity concentrations data in potatoes is 0.88. $^{137}$Cs activity concentrations data and models /4/ and /5/ are shown on Figure 2.

It should be noted that above models, showing mere functional dependence, are just simple tools for rough prediction of $^{137}$Cs activity concentration data in potatoes from fallout data and have no deeper physical meaning.

*The chain model for the transfer of $^{137}$Cs from fallout to potatoes*

The activity concentrations of $^{137}$Cs in potatoes have been related to the fallout deposition rates in a general regression model recommended by UNSCEAR (UNSCEAR, 1982):

$$A(i) = b_1 \dot{U}(i) + b_2 \dot{U}(i-1) + b_3 \sum_{m=0}^{\infty} e^{-\mu m} \dot{U}(i-m) \qquad /6/$$

where:

$A(i)$ is the activity concentration of observed radionuclide in food (potatoes) in the year $i$, the unit for $A(i)$ being Bq kg$^{-1}$,

$\dot{U}(i)$ the fallout deposition rate of radionuclide in year $i$ (Bq m$^{-2}$ y$^{-1}$),

$Ee^{-\mu m}$... the cumulative fallout deposit for radionuclide as the result of deposition in previous years (Bq m$^{-2}$ y$^{-1}$),

$\mu^{-1}$ the mean residence time of $^{137}$Cs in potatoes and,

$b_1, b_2, b_3$ the factors which are derived from reported data by regression analysis. The unit is Bq y kg$^{-1}$/(Bq m$^{-2}$).

The equation /6/ assumes the chain model for the transfer of radionuclides between environmental compartments, linking the input to the atmosphere to the dose in man, the successive compartments being atmosphere, Earth's surface, diet and tissue. The physical meaning of terms in model /6/ is as follows: the first term (*rate factor*) describes a direct deposition and transfer, the second term (*lag factor*) describes contamination through fallout from previous year, and the third term (*land factor*) reflects the contamination from fallout deposition accumulated from all preceding years, the exponential describing the combined physical decay and any other decrease in availability of considered radionuclide due to various other processes (like penetration in deeper soil layers, dilution etc.)

The $^{137}$Cs activity in potatoes and the fit obtained by equation /6/ are shown in Figure 3.

FIGURE 3 ABOUT HERE

The model prediction describes the real situation quite satisfactory; the coefficient of correlation between observed and modelled data being 0.93.

Regression analysis for the constant $\mu$ gives the value of 0.165 $y^{-1}$. The reciprocal value of the constant $\mu$, being 6.0 y, is the observed mean residence time of available $^{137}$Cs in potatoes and is just slightly lesser than the value predicted by equation /1/. Using equation /2/, effective ecological half-live is calculated to be 7.0 years.

The factors $b_1$, $b_2$, $b_3$ were estimated to be $1.26 \times 10^{-4}$, $9.00 \times 10^{-6}$ and $5.96 \times 10^{-5}$ respectively. The relative contribution of each of three factors to the modelled data of $^{137}$Cs activity concentration in potatoes obtained by equation /6/ varies from year to year, strongly depending upon the fallout deposition rate, the rate factor being the most important in the year in which the major contaminations occurred and the lag factor in the couple of following years. Therefore, in the events like the Chernobyl accident, the season of the year is the key factor influencing the radioactive contamination of crops and determining the transfer to man. Luckily, the Chernobyl accident happened at a time when direct contamination of leafy parts of potato plants was not so important.

As stated previously, for values of $A(t)$ and $U(t)$ assessed on the yearly basis the integration can be replaced by summation. By implementing that, the combination of equations /3/ and /6/ leads to another relation for the radioecological sensitivity:

$$P_{23} = \frac{b_1 + b_2 + b_3 e^{-\mu}}{1 - e^{-\mu}} \qquad /7/$$

Equation /7/ shows that radioecological sensitivity, i.e. the transfer coefficient from fallout to sample itself also consists of three factors. $P_{23}$ for $^{137}$Cs in potatoes utilizing equation /7/, i.e., after regression analysis, was estimated to be $4.66 \times 10^{-4}$ Bq y kg$^{-1}$ / Bq m$^{-2}$, which is just slightly lesser than the value of $P_{23}$ calculated directly by the equation /3/ after the summation of potatoes activity concentration and fallout deposition rates for the whole observed period.

By function minimization of experimental data on $^{137}$Cs in soil (Popović, 1966 - 1978; Bauman et al., 1979 - 1992; Kovač et al., 1993 - 1998; Marović et al., 1999 - 2004) to exponential function, it can be shown that the caesium migration to deeper soil layers is very slow process and that half-depth of radiocaesium penetration in uncultivated (i.e. undisturbed, non arable soils) is just few centimetres, which is well established fact (Szerbin et al., 1999; Isaksson et al., 2001). Therefore, most of the caesium activity is found in the first few centimetres of soil. In arable, i.e. cultivated soils, caesium consequently penetrates in deeper layers. Nevertheless, potatoes are not so much affected by caesium accumulated in soils, which results in relative minor contamination.

*Dosimetry*

Data on activity concentrations of $^{137}$Cs and $^{134}$Cs in potatoes allow for the estimate of the doses incurred by their consumption. Dose conversion factors, i.e. effective dose per unit intake via ingestion for the member of public older than 17 years are $1.3 \times 10^{-8}$ SvBq$^{-1}$ and $1.9 \times 10^{-8}$ SvBq$^{-1}$ for $^{137}$Cs and $^{134}$Cs respectively (International Atomic Energy Agency, 1996). As the ratio of dose conversion factors for $^{137}$Cs and $^{134}$Cs is $\approx 0.7$, it implies that ingestion of $^{134}$Cs contributes about 30% more to the dose, compared to ingestion of the same activity concentration of $^{137}$Cs.

Collective effective dose incurred due to food consumption over certain time period, depends on the activity of a radionuclide and on the quantity of food that is consumed. The dose can be expressed as:

$$E = C \sum_k D_k^{cf} A_k \qquad /8/$$

where:

$E$      is the effective dose in Sv,

*C*     total annual *per caput* consumption of potatoes (kg),

$D_k^{cf}$     dose conversion factor for radionuclide *k*, i.e. effective dose per unit input, which converts the ingested activity to effective dose (SvBq$^{-1}$) and

$A_k$     mean yearly specific activity of radionuclide *k* in food (Bqkg$^{-1}$).

Based on the statistical data for annual consumption rate for potatoes of 40 kg, the estimated total effective dose due to $^{137}$Cs and $^{134}$Cs ingestion by potato consumption for the Croatian population and overall observed period, i.e. 1986-2005 is 13.2 person-Sv. Out of this, 9.6 person-Sv could be attributed to $^{137}$Cs and 3.6 person-Sv to $^{134}$Cs. 4.2 person-Sv refers to the year 1986 (2.6 person-Sv and 1.6 person-Sv for $^{137}$Cs and $^{134}$Cs respectively).

The annual effective collective doses are shown on Figure 4.

FIGURE 4 ABOUT HERE

It should be noted that these doses are rather small since for the overall 1986-2005 period per caput dose is . 2.9 : Sv. Consequently, it can be argued that after the Chernobyl accident consumption of potatoes was not the critical pathway for human intake of radiocaesium from the environment in Croatia.

**CONCLUSIONS**

Long-term investigations of radiocaesium activity concentrations in potatoes and fallout in the Republic of Croatia showed that the transfer of radiocaesium from fallout to potatoes and finally to humans was pretty low, which has been observed also elsewhere (Aarkrog, 1994). The $^{137}$Cs transfer coefficient from fallout to potatoes was of the same order of magnitude (10$^{-4}$) as it was found to be for the other foodstuffs (10$^{-4}$ - 10$^{-3}$). For the events like the Chernobyl nuclear accident the season of the year is the most important factor determining the transfer of radioactive contamination from potatoes to men.

The effective mean residence time of $^{137}$Cs in potatoes, describing the decrease in the observed $^{137}$Cs activity concentration with time, was estimated to be between 6 and 7 years.

Generally, a few years after the Chernobyl nuclear accident the activities of fission radionuclides $^{137}$Cs and $^{134}$Cs in potatoes in Croatia were relatively low, in magnitude comparable to the pre-Chernobyl period. Consequently the doses to general population received by radiocaesium in potatoes are small, in spite of considerably large consumption of potatoes in Croatia.

ACKNOWLEDGEMENTS

The authors wish to thank Mrs. Ljerka Petroci for her excellent technical assistance and help.

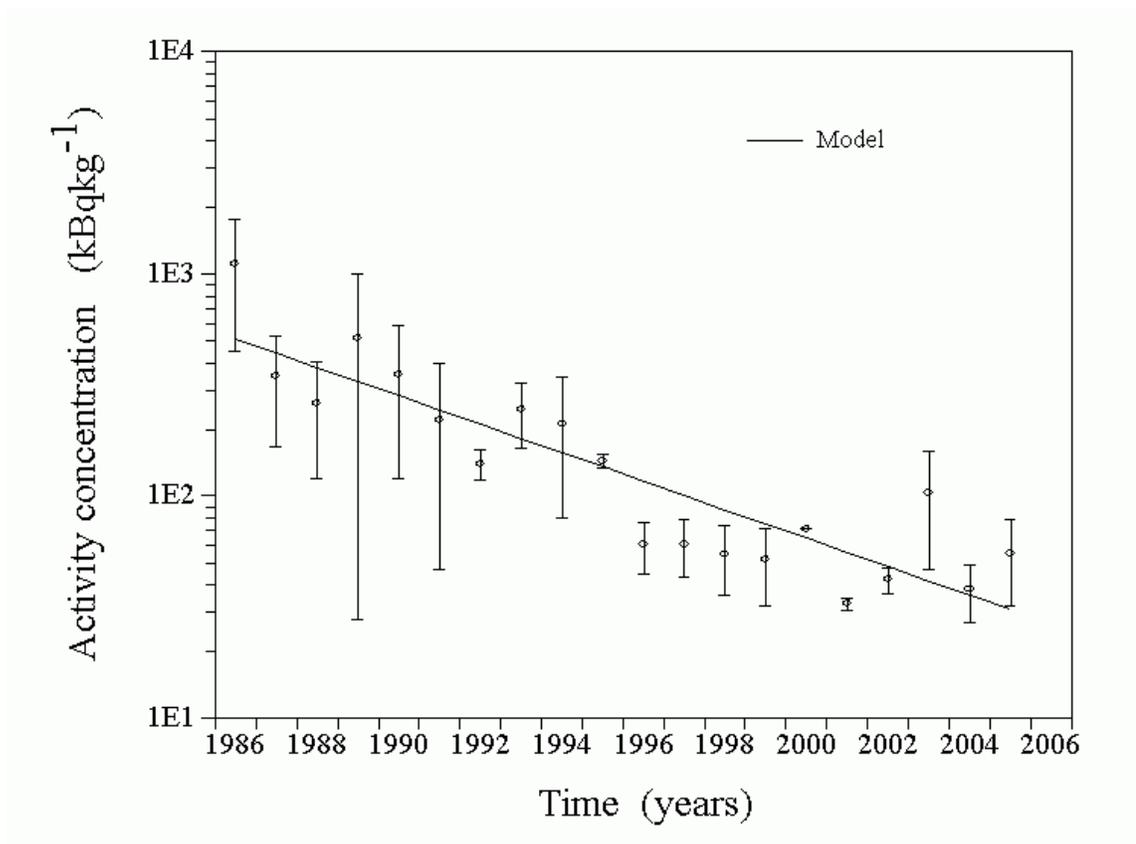

Figure 1

Observed and modelled $^{137}$Cs activity concentrations in potatoes using exponential function given by equation /1/ as a function of time

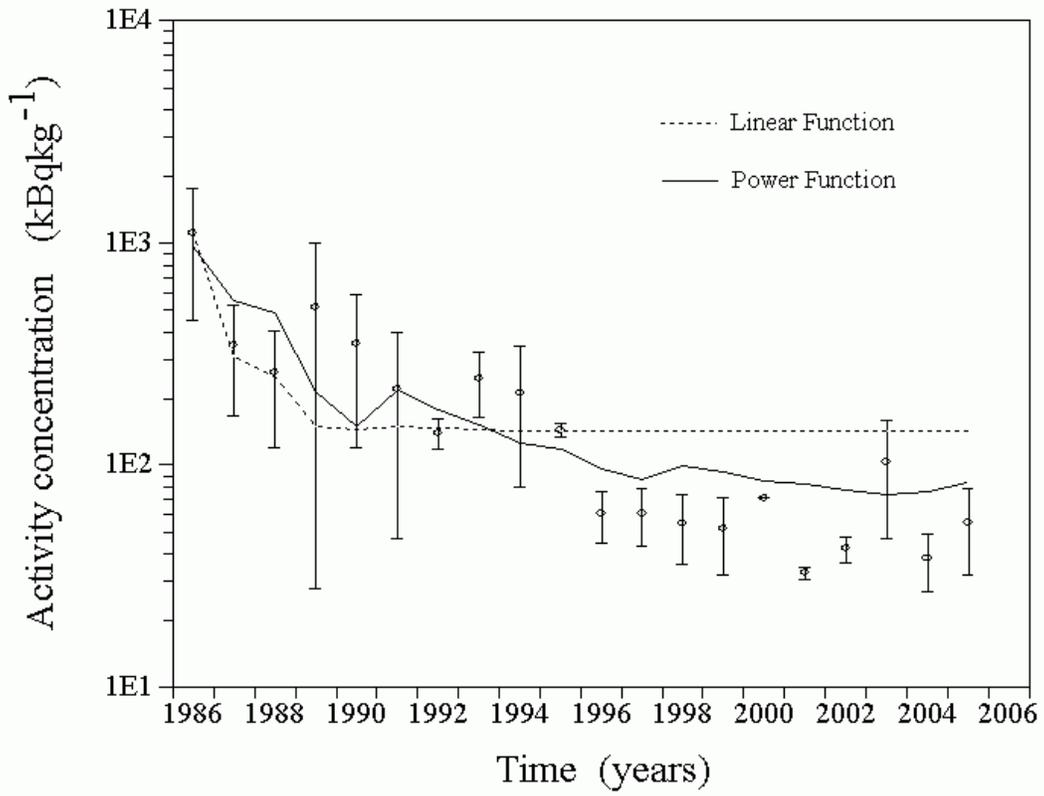

Figure 2

Observed and modelled $^{137}$Cs activity concentrations in potatoes using linear and power fit given by equations /4/ and /5/ as a function of time

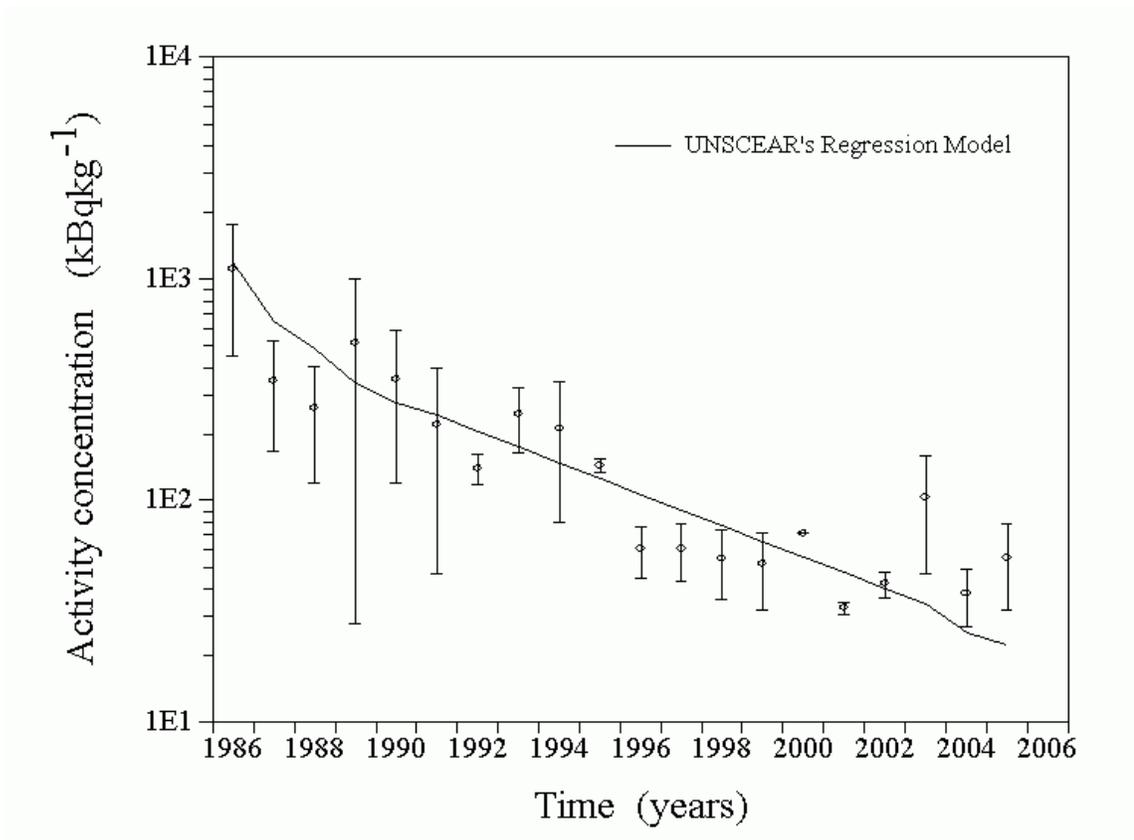

Figure 3

Observed and modelled $^{137}$Cs activity concentrations in potatoes using UNSCEAR's regression model given by equation /6/ as a function of time

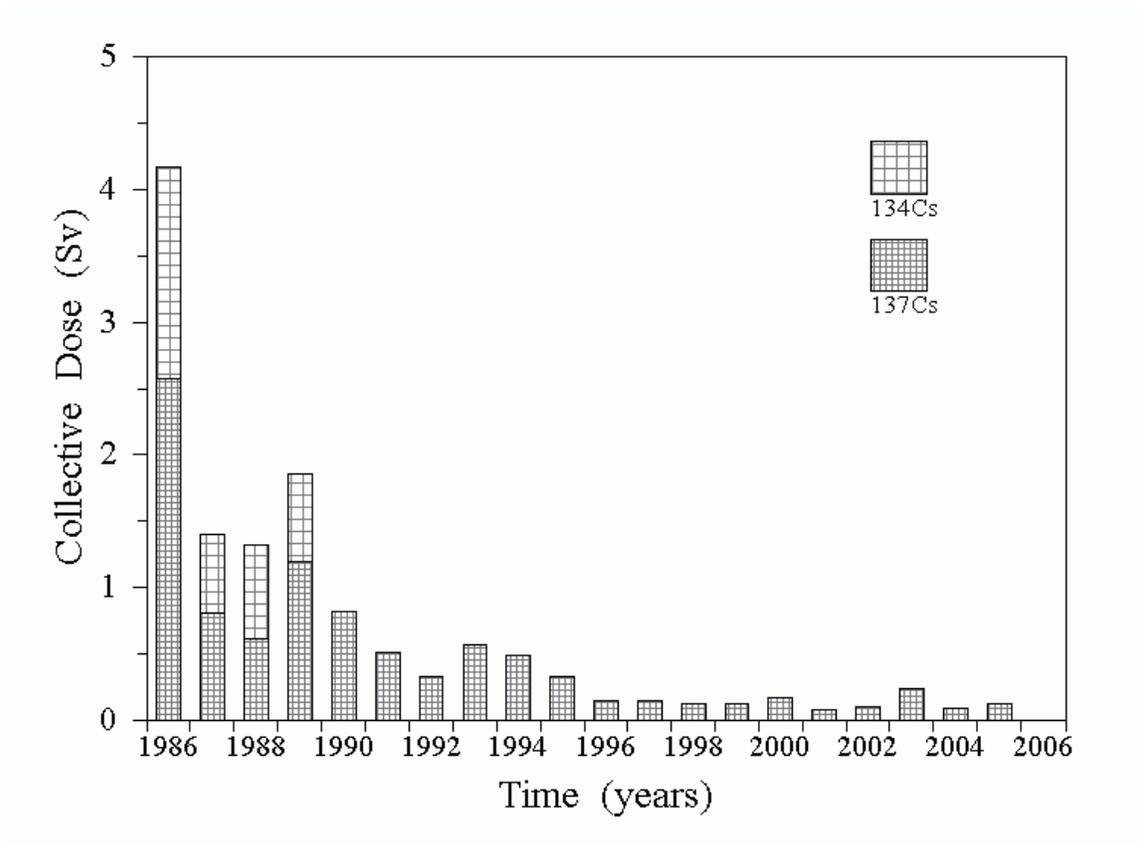

Figure 4

Estimated annual collective effective doses for Croatian population received by $^{134}$Cs and $^{137}$Cs intake due to consumption of potatoes